\documentstyle[prl,aps,epsf]{revtex}

\newcommand{\bea}{\begin{eqnarray}}
\newcommand{\eea}{\end{eqnarray}}
\newcommand{\beq}{\begin{equation}}
\newcommand{\eeq}{\end{equation}}
\newcommand{\del}{\partial}
\newcommand{\lishi}{\langle\!\langle}
\newcommand{\rishi}{\rangle\!\rangle}

\begin{document}

\draft

\twocolumn[\hsize\textwidth\columnwidth\hsize\csname
@twocolumnfalse\endcsname


\title{Coulomb-gas approach for boundary conformal field theory}

\author{Shinsuke Kawai\footnotemark[1]}
\address{Theoretical Physics, Department of Physics, 
 University of Oxford, \\1 Keble Road, Oxford OX1 3NP, United Kingdom}
\date{\today}
\maketitle

\begin{abstract}
We present a construction of boundary states based on
the Coulomb-gas formalism of Dotsenko and Fateev.
It is shown that Neumann-like coherent states on the charged bosonic 
Fock space provide a set of boundary states with consistent modular
properties.
Such coherent states are characterised by the boundary charges, 
which are related to the number of bulk screening operators through the 
charge neutrality condition.
We illustrate this using the Ising model as an example, 
and show that all of its known consistent boundary states are reproduced in 
our formalism.
This method applies to $c<1$ minimal conformal theories and provides
an unified computational tool for studying boundary states of such theories.
\end{abstract}

\noindent
\preprint{
OUTP0203P, hep-th/0201146
}

\vskip2pc]
\footnotetext[1]{E-mail address: \tt s.kawai1@physics.ox.ac.uk}

\section{Introduction}

The basic concepts and techniques of boundary conformal field theory (BCFT) 
were introduced in the eighties. 
Much of the seminal work was done by Cardy, who discussed surface
critical behaviour and invented a method to calculate boundary correlation 
functions\cite{cardy1}, studied the restriction on the operator content
imposed by boundary conditions\cite{cardy2}, developed a systematic 
classification of boundary states based on the modular transformation and 
introduced the concept of boundary operators\cite{cardy3,cardylewellen}.
After the relatively dormant era of the middle nineties, this field is now 
enjoying its second stage of development. 
The growing interest in BCFT is largely motivated by noticing the importance 
of boundary states of open strings after the discovery of D-branes.
Classification of boundary states is now recognised as of prime
importance since D-branes are an essential element of non-perturbative
string theory.

Our understanding of the underlying algebraic structure of BCFT has recently
been improved enormously.
Importance of complete sets of boundary conditions\cite{pradisi} has been
recognised widely. 
An extra boundary condition was discovered in the simplest non-diagonal 
minimal model\cite{3sp1} and the resulting set of boundary conditions was
shown to be complete\cite{3sp2}.
The sewing relations originated by Cardy and 
Lewellen\cite{cardylewellen,lewellen} were explicitly solved for some 
cases\cite{runkel}, and it is now understood that solving Cardy's consistency 
condition reduces to finding non-negative integer-valued matrix 
representations of the Verlinde algebra\cite{bppz}.
A rational conformal theory is rational with respect to a symmetry which is
in general larger than the Virasoro symmetry, whereas boundary states only
need to be invariant under the Virasoro symmetry.
Classification of boundary states from such a symmetry-breaking viewpoint
is also being done\cite{fsch,bfsch}.
The algebraic construction of boundary states (or D-branes) is now being 
extended to various rational conformal theories far beyond the minimal models.

In this paper we would like to consider another approach for BCFT, namely,
the construction of boundary states from free fields. 
This is particularly important from the practical point of view since any
correlation function should be calculable ab initio from the operator algebra. 
Such BCFTs for free bosons and fermions have been well established for a long 
time, and they are indeed essential building blocks of the open string theory.
Aside from bosons and fermions, the boundary states of symplectic fermions at 
$c=-2$ were constructed recently\cite{kawaiwheater}. 
However, to the author's knowledge, such an approach for other CFTs seems to 
be absent.
In order to generalise the free-field construction of boundary states we
re-formulate the Coulomb-gas picture of Dotsenko and Fateev\cite{dotfat} in 
the presence of a boundary.
Work in this direction was done by Schulze\cite{schulze}, who discussed
Coulomb-gas system on the half plane and calculated boundary correlation
functions using contour integrals.
In the present paper we shall consider the system on an annulus which is a
suitable arena for discussing modular properties and algebraic structures of 
boundary states.
The goal of this paper is to reproduce the result of $c<1$ diagonal minimal 
conformal theories\cite{cardy3} by explicit construction of boundary states 
in Fock space representation.

The plan of this paper is as follows. We start in the next section
by reviewing the Coulomb-gas formalism on the Riemann surfaces and fix our
notation.
In Sec. III the charged bosonic Fock space (CBFS) is defined for
the theory on an annulus. 
We also construct the boundary coherent states on CBFS and find conditions 
for the conformal invariance of such states.
The charge-neutrality conditions for the boundary Coulomb-gas are considered
and the closed-string channel amplitudes are calculated in Sec. IV. 
We illustrate our method in Sec V using the Ising model as an example.
In Sec. VI we summarise and conclude. 


\section{Coulomb-gas and the charged bosonic Fock space}

The essential ingredient of the Coulomb-gas formalism is the non-minimal
coupling of the free scalar field to the background curvature.
This makes the $U(1)$ symmetry anomalous, modifying the central charge and 
the conformal dimensions of $c=1$ theory to generate the minimal models. 
In this section we collect the basic components of the Coulomb-gas 
formalism without the boundary\cite{dotfat,felder,ketov,dFMS}.
Variation of the action
\beq
{\cal S}=\frac{1}{8\pi}\int d^2x\sqrt g(\del_\mu\Phi\del^\mu\Phi
+2\sqrt 2\alpha_0i\Phi R),
\eeq
with respect to the metric gives the energy-momentum tensor
\beq
T(z)=-2\pi T_{zz}
=-\frac 12 :\del\varphi\del\varphi:+i\sqrt 2\alpha_0\del^2\varphi,
\eeq
where $\varphi$ is the holomorphic part of the boson,
$\Phi(z,\bar z)=\varphi(z)+\bar\varphi(\bar z)$. 
The antiholomorphic part is similar. 
From $T(z)$ the central charge is read off as
\beq
c=1-24\alpha_0^2.
\label{eqn:central}
\eeq
The {\em chiral} vertex operator defined as
\beq
V_\alpha(z)=:e^{i\sqrt 2\alpha\varphi(z)}:
\eeq
then has the conformal dimension
$
h_\alpha=\alpha^2-2\alpha_0\alpha,
$
which is easily verified by computing the OPE with $T(z)$. 
Among these vertex operators,
$V_{\pm}(z)\equiv V_{\alpha_{\pm}}(z)$ with 
$\alpha_{\pm}=\alpha_0\pm\sqrt{\alpha_0^2+1}$
play a special role. 
They have conformal dimensions $1$ and the closed contour integrals,
\beq
Q_\pm\equiv\oint dz V_\pm(z),
\label{eqn:screening}
\eeq
are the screening operators which do not change the conformal properties but
carry charges. 
The condition that the fields must be screened by such screening operators
leads to the quantisation of the spectrum,
\beq
\alpha_{r,s}=\frac 12(1-r)\alpha_++\frac 12(1-s)\alpha_-,
\eeq
where $r$ and $s$ are positive integers.
The vertex operators $V_{\alpha_{r,s}}(z)$ then have conformal dimensions
\beq
h_{r,s}=\frac 14(r\alpha_++s\alpha_-)^2-\alpha_0^2,
\eeq
and are identified with the operators appearing in the Kac formula.

The Hilbert space of the theory defined on a Riemann surface is a
direct sum of charged bosonic Fock spaces (CBFSs) with BRST 
projection\cite{felder}. 
The chiral CBFS $F_{\alpha, \alpha_0}$ with vacuum charge $\alpha$ and 
background charge $\alpha_0$ is built on the highest-weight vector
$\vert\alpha;\alpha_0\rangle$ as a representation of the Heisenberg algebra
\beq
[a_m,a_n]=m\delta_{m+n,0},
\label{eqn:heisenberg}
\eeq
where $a_n$ are the mode operators defined by
\beq
\varphi(z)=\varphi_0-ia_0\ln z+i\sum_{n\neq 0}\frac{a_n}{n}z^{-n}.
\label{eqn:phimode}
\eeq
The zero-mode operators satisfy the commutation relation
$[\varphi_0,a_0]=i$. The highest-weight vector is constructed from the vacuum
$\vert 0;\alpha_0\rangle$ by operating with $e^{i\sqrt 2 \alpha\varphi_0}$,
\beq
\vert\alpha;\alpha_0\rangle=e^{i\sqrt 2 \alpha\varphi_0}
\vert 0;\alpha_0\rangle,
\label{eqn:hwvin}
\eeq
and is annihilated by the action of $a_{n>0}$. The charge $\alpha$ is related 
to the eigenvalue of $a_0$ by
\beq
a_0\vert\alpha;\alpha_0\rangle=\sqrt 2\alpha\vert\alpha;\alpha_0\rangle.
\label{eqn:a0in}
\eeq 
The Virasoro generators are written in terms of the mode operators as
\bea
&&L_{n\neq 0}=\frac 12\sum_{k\in Z}a_{n-k}a_k-\sqrt 2\alpha_0(n+1)a_n,
\label{eqn:virasoro1}\\
&&L_0=\sum_{k\geq1}a_{-k}a_k+\frac 12 a_0^2-\sqrt 2\alpha_0a_0.
\label{eqn:virasoro2}
\eea
With these generators the CBFS $F_{\alpha,\alpha_0}$ has the structure of 
a Virasoro module. It is easy to check that
\beq
L_0\vert\alpha;\alpha_0\rangle=(\alpha^2-2\alpha\alpha_0)\vert\alpha;
\alpha_0\rangle,
\eeq
that is, the conformal dimension of $\vert\alpha;\alpha_0\rangle$
is $\alpha^2-2\alpha\alpha_0$. 
Because of $[L_0, a_{-n}]=na_{-n}\;(\forall n\geq 0)$, 
$F_{\alpha,\alpha_0}$ is graded by $L_0$ and written as
\beq
F_{\alpha,\alpha_0}=\bigoplus_{n=0}^\infty(F_{\alpha,\alpha_0})_n,
\eeq
where
$(F_{\alpha,\alpha_0})_n$ is the subspace with conformal dimension
$\alpha^2-2\alpha\alpha_0+n$. 
Counting the number of states the character of $F_{\alpha,\alpha_0}$ is
found to be
\beq
\chi_{\alpha,\alpha_0}(q)\equiv\mathop{\rm Tr}_{F_{\alpha,\alpha_0}}
q^{L_0-c/24}
=\frac{q^{(\alpha-\alpha_0)^2}}{\eta(\tau)},
\label{eqn:cbfschar}
\eeq
where $q=e^{2\pi i\tau}$ is the modular parameter and 
$\eta(\tau)\equiv q^{1/24}\prod_{n\geq1}(1-q^n)$ is the Dedekind eta function.

The dual space $F_{\alpha,\alpha_0}^*$ of $F_{\alpha,\alpha_0}$ is built on a
contravariant highest-weight vector $\langle\alpha;\alpha_0\vert$
satisfying the condition
\beq
\langle\alpha;\alpha_0\vert\alpha;\alpha_0\rangle=\kappa,
\label{eqn:norm}
\eeq
where $\kappa$ is a normalisation factor which is usually set to $1$ in
unitary models.
The modules are endowed with a dual Virasoro structure
\beq
\langle\omega\vert L_{-n}\xi\rangle=\langle\omega L_n\vert\xi\rangle
\eeq
for any $\langle\omega\vert\in F_{\alpha,\alpha_0}^*$, 
$\vert\xi\rangle\in F_{\alpha,\alpha_0}$.
This dual structure naturally incorporates the transpose $A^t$ of an operator
$A$ through the relation
\beq
\langle\omega\vert A\xi\rangle=\langle\omega A^t\vert\xi\rangle.
\label{eqn:transpose}
\eeq
In particular, 
$L^t_{-n}=L_{n}$, $a^t_{-n}=2\sqrt 2\alpha_0\delta_{n,0}-a_n$.
With this definition of transpose, $F_{\alpha,\alpha_0}^*$ is shown to be 
a Fock space isomorphic to $F_{2\alpha_0-\alpha,\alpha_0}$.
The contravariant highest-weight vector $\langle\alpha;\alpha_0\vert$
is annihilated by the action of $a_n$ for $n<0$ (or $a_n^t$ for $n>0$),
\beq
\langle\alpha;\alpha_0\vert a_{n<0}=0.
\eeq
From the uniqueness of the expression
$\langle\alpha;\alpha_0\vert a_0\vert\alpha;\alpha_0\rangle$ 
and the right operation of the zero mode (\ref{eqn:a0in}) we immediately have
\beq
\langle\alpha;\alpha_0\vert a_0 =\sqrt 2\alpha\langle\alpha;\alpha_0\vert.
\label{eqn:a0out}
\eeq
Analogously to (\ref{eqn:hwvin}) we find
\beq
\langle\alpha;\alpha_0\vert
=\langle 0;\alpha_0\vert e^{-i\sqrt 2\alpha\varphi_0},
\label{eqn:hwvout}
\eeq
where the contravariant vector $\langle 0;\alpha_0\vert$ is the vacuum 
with the normalisation $\langle 0;\alpha_0\vert 0;\alpha_0\rangle=\kappa$.
From (\ref{eqn:hwvin}) and (\ref{eqn:hwvout}), 
the in-state $\vert\alpha;\alpha_0\rangle$ and the out-state
$\langle\alpha;\alpha_0\vert$ are interpreted as possessing charges 
$\alpha$ and $-\alpha$, respectively.
The non-vanishing inner product (\ref{eqn:norm}) is consistent with the 
neutrality of the total charge, $-\alpha+\alpha=0$. 
Since the inner product must vanish when the total charge is not zero,
we have in general
\beq
\langle\alpha;\alpha_0\vert\beta;\alpha_0\rangle=\kappa\delta_{\alpha,\beta}.
\label{eqn:shapovalov}
\eeq 

On the plane the minimal conformal theory is realized through the usual
radial quantisation scheme, by sending the in-state to zero and the out-state
to infinity. Expectation values are usually taken between
$\langle 2\alpha_0;\alpha_0\vert$ and $\vert 0;\alpha_0\rangle$,
which is interpreted as placing a charge $-2\alpha_0$ at infinity.
Correlation functions of vertex operators are calculated with suitable
insertion of the screening operators to realise the charge neutrality,
leading in general to integral representations.
The Coulomb-gas formalism also applies to Riemann surfaces of higher genus
and such theories have been studied by many 
authors\cite{felder,FelSil,DiPFHLS,flms}.
On the torus it is shown that taking the trace over the BRST cohomology space
is equivalent to the alternated summation\cite{felder}. 
For example, the zero-point function on the torus for the conformal block 
corresponding to the representation $(r,s)$ of the minimal models is 
calculated in Coulomb-gas method as\cite{felder}
\beq
{\rm Tr}_{(r,s)}q^{L_0-c/24}=\frac{1}{\eta(\tau)}
(\Theta_{pr-p's,pp'}(\tau)-\Theta_{pr+p's,pp'}(\tau)),
\label{eqn:char}
\eeq
which is nothing but the Rocha-Caridi character formula\cite{rocha}
as it should be.
Here, we have defined the Jacobi theta function as
$
\Theta_{\lambda,\mu}(\tau)\equiv\sum_{k\in Z}q^{(2\mu k+\lambda)^2/4\mu}
$.
Formulae for Jacobi theta and Dedekind eta functions are summarised in 
App. A.


\section{CBFS with boundary}

In this section we discuss the Fock space representation of BCFT where the 
interplay between holomorphic and antiholomorphic sectors is
important.
Let us start with the geometry of the upper half-plane.
We define $\zeta=x+iy$, $x$, $y\in R$ and consider a CFT defined on the region
${\rm Im} \zeta\geq 0$. 
The boundary is $y=0$, or $\zeta=\bar\zeta$.
The antiholomorphic dependence of the correlators on the upper half plane may 
be mapped into the holomorphic dependence on the lower half plane\cite{cardy1}.
This introduces a mirror image on the lower half plane, and the boundary 
condition tells how the images on the upper and lower half-planes are glued
on the mirror, $\zeta=\bar\zeta$.
The energy-momentum tensor on the lower half plane is obtained by the
mapping from the upper half plane, $T(\zeta^*)=\bar T(\bar\zeta)$. 
The condition on the boundary
\beq
\left[T(\zeta)-\bar T(\bar\zeta)\right]_{\zeta=\bar\zeta}=0,
\label{eqn:Tcondzeta}
\eeq
indicates the absence of the energy-momentum flow across the boundary.
Since the energy-momentum tensor is the generator of conformal transformations,
(\ref{eqn:Tcondzeta}) also means the conformal invariance of the boundary.
Going from the upper half plane (or holomorphic part) to the lower half plane 
(antiholomorphic part) is generally accompanied by a {\em parity} 
transformation ${\cal P}$.
The free boson transforms under ${\cal P}$ as
$\varphi(\zeta)\rightarrow\Omega\bar\varphi(\bar\zeta)$,
$\Omega=\pm 1$.
This leads to the condition on the boundary
\beq
\left[\varphi(\zeta)-\Omega\bar\varphi(\bar\zeta)\right]_{\zeta=\bar\zeta}=0.
\label{eqn:phicondzeta}
\eeq 
When $\Omega=1$, the non-chiral free boson 
$\Phi(\zeta,\bar\zeta)=\varphi(\zeta)+\bar\varphi(\bar\zeta)$ is
a scalar and the boundary condition is called Neumann, 
whereas when $\Omega=-1$, $\Phi(\zeta,\bar\zeta)$ is a pseudo-scalar and such 
boundary condition is called Dirichlet.
Under the parity transformation the chiral vertex operators 
$V_\alpha(\zeta)=:e^{i\sqrt 2\alpha\varphi(\zeta)}:$ are mapped into
$\bar V_\alpha(\bar\zeta)= :e^{i\sqrt 2\alpha\Omega\bar\varphi(\bar\zeta)}:$.
When $\Omega=-1$ (Dirichlet) the mirror image has a charge 
$\Omega\alpha=-\alpha$ which has the opposite sign from the original one. 
In the Neumann case ($\Omega=1$), the mirror and the original vertex operators
have the same charge $\alpha$.
Coulomb-gas system on the half plane was studied in \cite{schulze},
where the boundary correlation functions of the Ising model are calculated 
using the mirroring technique of \cite{cardy1}. 

In this paper we mainly study BCFT defined on a finite cylinder,
or an annulus.
We consider a finite cylinder of length $T$ and circumference $L$, or an 
annulus on the $z$-plane with $1\leq\vert z\vert\leq\exp(2\pi T/L)$. 
We also introduce a modular parameter as $\tilde q=e^{2\pi i\tilde\tau}$, 
$\tilde\tau=2iT/L$. 
With this the annulus is $1\leq\vert z\vert\leq\tilde q^{-1/2}$.
We regard this cylinder as a propagating closed string, and call the direction
along it as {\em time}.
A merit of considering such a geometry is that the familiar energy-momentum 
tensor for the full-plane may be used without modification.
We conformally map a semi-annular domain in the upper-half $\zeta$-plane
onto a full-annulus in the $z$-plane by $z=\exp(-2\pi iw/L)$ and
$w=(T/\pi)\ln\zeta$. The boundary $\zeta=\bar\zeta$ is then mapped on the
$z$-plane to $\vert z\vert=1$, $\exp(2\pi T/L)$.
Since the $z$-plane allows radial quantization, the conformal invariance
(\ref{eqn:Tcondzeta}) on the $\vert z\vert=1$ boundary becomes the conditions 
on the quantum states $\vert B\rangle$\cite{cardy3,ishibashi},
\beq
(L_k-\bar L_{-k})\vert B\rangle=0.
\label{eqn:Tcondz}
\eeq
As $\varphi(\zeta)$ and $\bar\varphi(\bar\zeta)$ are not primary, 
the condition (\ref{eqn:phicondzeta}) cannot be mapped to the annulus.
However, the derivative of {\em uncharged} bosons are primary and 
\beq
\left[\del\varphi(\zeta)-\Omega\bar\del\bar\varphi(\bar\zeta)\right]_
{\zeta=\bar\zeta}=0
\label{eqn:dphicondzeta}
\eeq 
on the $\zeta$-plane is mapped on the $z$-plane as
\beq
(a_n+\Omega\bar a_{-n})\vert B\rangle=0.
\label{eqn:dphicondz}
\eeq
This expression no longer makes sense for the {\em charged} 
bosons since $\del\varphi$ and $\bar\del\bar\varphi$ cease to be primary when 
they are couple to the background curvature.
However, (\ref{eqn:Tcondz}) is still valid and is indeed a necessary condition
for the conformally invariant boundary states.
The vertex operators are safely mapped to $z$-plane since they remain
primary.
In the rest of this section we construct a Fock space representation of 
boundary states which satisfy the conformal invariance condition
(\ref{eqn:Tcondz}). 

Our starting point is recalling that a BCFT consists of a pair of
chiral CFTs whose holomorphic and antiholomorphic sectors are glued together 
on the boundary.
The construction of the boundary states then requires a Fock space which is
common to both holomorphic and antiholomorphic sectors. 
As we have the same central charge $c$ for both holomorphic and antiholomorphic
sectors, $\alpha_0$, which is related to $c$ by (\ref{eqn:central}), is common 
to both sectors, although we are free to choose different vacuum charges 
for each sector. 
Hence let us define the highest-weight vectors at the two boundaries of the
annulus as
$\vert\alpha,\bar\alpha;\alpha_0\rangle$ and 
$\langle\alpha,\bar\alpha;\alpha_0\vert$, satisfying
\bea
&&a_0\vert\alpha,\bar\alpha;\alpha_0\rangle
=\sqrt 2\alpha\vert\alpha,\bar\alpha;\alpha_0\rangle,
\label{eqn:zeromodeop1}\\
&&\bar a_0\vert\alpha,\bar\alpha;\alpha_0\rangle
=\sqrt 2\bar\alpha\vert\alpha,\bar\alpha;\alpha_0\rangle,
\label{eqn:zeromodeop2}\\
&&\langle\alpha,\bar\alpha;\alpha_0\vert a_0
=\langle\alpha,\bar\alpha;\alpha_0\vert\sqrt 2\alpha,
\label{eqn:zeromodeop3}\\
&&\langle\alpha,\bar\alpha;\alpha_0\vert\bar a_0
=\langle\alpha,\bar\alpha;\alpha_0\vert\sqrt 2\bar\alpha,
\label{eqn:zeromodeop4}
\eea
which are essentially the direct products of holomorphic and antiholomorphic
parts of (\ref{eqn:a0in}), (\ref{eqn:a0out}).
The state $\vert\alpha,\bar\alpha;\alpha_0\rangle$ has {\em holomorphic}
charge $\alpha$ and {\em antiholomorphic} charge $\bar\alpha$, and
$\langle\alpha,\bar\alpha;\alpha_0\vert$ has holomorphic charge $-\alpha$ 
and antiholomorphic charge $-\bar\alpha$.
The mode operators of the antiholomorphic sector are defined, similarly to the 
holomorphic part (\ref{eqn:phimode}), by the mode expansion of
$\bar\varphi(\bar z)$ as
\beq
\bar\varphi(\bar z)=\bar\varphi_0-i\bar a_0\ln\bar z
+i\sum_{n\neq 0}\frac{\bar a_n}{n}\bar z^{-n}.
\label{eqn:barphimode}
\eeq
The antiholomorphic mode operators satisfy the same Heisenberg algebra as their
holomorphic counterpart:
\bea
&&[\bar a_m,\bar a_n]=m\delta_{m+n,0},
\label{eqn:anti-heisenberg}\\
&&[\bar\varphi_0,\bar a_0]=i.
\eea
There is a subtlety in the treatment of $\bar\varphi_0$ and 
$\bar a_0$ since the zero mode of the boson $\Phi(z,\bar z)$ does not naturally
decouple into left and right.
We split them into two identical and independent copies such that  
$[\varphi_0, \bar a_0]=[\bar\varphi_0, a_0]=0$.
In such decomposition the existence of the dual field is 
implicit\cite{koganwheater}.
The highest-weight vector $\vert\alpha,\bar\alpha;\alpha_0\rangle$
is annihilated by the action of $a_{n>0}$ and $\bar a_{n>0}$, and the
contravariant highest-weight vector 
$\langle\alpha,\bar\alpha;\alpha_0\vert$ is annihilated by 
$a_{n<0}$ and $\bar a_{n<0}$.
Following (\ref{eqn:shapovalov})
we assume the highest-weight vectors are normalised as
\beq
\langle\alpha,\bar\alpha;\alpha_0\vert\beta,\bar\beta;\alpha_0\rangle
=\kappa'\delta_{\alpha,\beta}\delta_{\bar\alpha,\bar\beta},
\label{eqn:hwvnorm}
\eeq
where $\kappa'$ is a normalisation factor, which may be set to $1$ if the 
sector is unitary. If $\kappa'$ is negative we set it to $-1$.

We are looking for conformally invariant boundary states built
on the highest-weight vectors 
$\vert\alpha,\bar\alpha;\alpha_0\rangle$ and 
$\langle\alpha,\bar\alpha;\alpha_0\vert$.
Since we know that such states for (uncharged) bosonic strings are found in 
the form of coherent states in string theory, let us start with an ansatz 
\bea
\vert B_{\alpha,\bar\alpha;\alpha_0}\rangle_\Omega
=\prod_{k>0}\exp\left(-\frac{\Omega}{k}a_{-k}\bar a_{-k}\right)
\vert\alpha,\bar\alpha;\alpha_0\rangle,
\label{eqn:bket}\\
{}_\Omega\langle B_{\alpha,\bar\alpha;\alpha_0}\vert
=\langle\alpha,\bar\alpha;\alpha_0\vert\prod_{k>0}
\exp\left(-\frac{1}{k\Omega}a_k\bar a_k\right).
\label{eqn:bbra}
\eea
These states satisfy
\bea
(a_n+\Omega\bar a_{-n})\vert B_{\alpha,\bar\alpha;\alpha_0}\rangle_\Omega=0\;\;
(n\neq 0),\\
{}_\Omega\langle B_{\alpha,\bar\alpha;\alpha_0}\vert(a_n+\Omega\bar a_{-n})=0\;\;
(n\neq 0).
\eea
Using the expression of Virasoro operators (\ref{eqn:virasoro1}) 
(\ref{eqn:virasoro2}) we see that
$\vert B_{\alpha,\bar\alpha;\alpha_0}\rangle_\Omega$ does {\em not} satisfy the
condition (\ref{eqn:Tcondz}) straightaway. For example, we have
\bea
&&(L_n-\bar L_{-n})\vert B_{\alpha,\bar\alpha;\alpha_0}\rangle_\Omega
\nonumber\\
&&=\prod_{k>0}\exp\left(-\frac{\Omega}{k}a_{-k}\bar a_{-k}\right)\nonumber\\
&&\times\left\{\sqrt 2 \bar a_{-n}
[(\Omega-1)n\alpha_0+(\Omega+1)\alpha_0-\Omega\alpha-\bar\alpha]\right.\nonumber\\
&&+\left.\frac 12\sum_{0<j<n}\bar a_{-j}\bar a_{j-n}(\Omega^2-1)\right\}
\vert\alpha,\bar\alpha;\alpha_0\rangle
\label{eqn:Lncond}
\eea
for $n>0$, and 
\bea
&&(L_0-\bar L_0)\vert B_{\alpha,\bar\alpha;\alpha_0}\rangle_\Omega
\nonumber\\
&&=\prod_{k>0}\exp\left(-\frac{\Omega}{k}a_{-k}\bar a_{-k}\right)\nonumber\\
&&\times\left\{(\alpha-\bar\alpha)(\alpha+\bar\alpha-2\alpha_0)\right\}
\vert\alpha,\bar\alpha;\alpha_0\rangle,
\label{eqn:L0cond}
\eea
which are in general not zero.
However, it can be easily seen that the expressions (\ref{eqn:Lncond}) and 
(\ref{eqn:L0cond}) do vanish when 
\beq
\Omega=1,
\label{eqn:omega}
\eeq
and 
\beq
\alpha+\bar\alpha-2\alpha_0=0,
\label{eqn:alpha}
\eeq
even for $\alpha_0\neq 0$. 
It is easily verified that these conditions also lead to 
$(L_n-\bar L_{-n})\vert B_{\alpha,\bar\alpha;\alpha_0}\rangle_\Omega=0$ 
for $n<0$
and are indeed a sufficient condition for the conformal invariance.
Similarly it can be checked that
${}_\Omega\langle B_{\alpha,\bar\alpha;\alpha_0}\vert(L_n-\bar L_{-n})=0$
as long as $\Omega=1$ and $\alpha+\bar\alpha-2\alpha_0=0$.
Note that the ``Dirichlet'' condition $\Omega=-1$ is not compatible with the 
conformal invariance for non-zero $\alpha_0$ because of the term proportional 
to $n$ in (\ref{eqn:Lncond}).
In the rest of this paper we shall consider the conformally invariant 
boundary states satisfying the conditions 
(\ref{eqn:omega}) and (\ref{eqn:alpha}). 
Since the antiholomorphic charge is determined by the condition 
(\ref{eqn:alpha}), such boundary states are characterised by only one parameter
$\alpha$, apart from the value of the background charge $\alpha_0$ which 
is fixed by the central charge. 
For simplicity we shall denote these boundary states as
\beq
\vert B(\alpha)\rangle
=\vert B_{\alpha,2\alpha_0-\alpha;\alpha_0}\rangle_{\Omega=1},
\label{eqn:cohbsket}
\eeq
and
\beq
\langle B(\alpha)\vert
={}_{\Omega=1}\langle B_{\alpha,2\alpha_0-\alpha;\alpha_0}\vert.
\label{eqn:cohbsbra}
\eeq
The background charge $\alpha_0$ is suppressed since no confusion arises.


\section{Coherent and consistent boundary states}

Identifying boundary states which may be realised in a physical system
is one of the main goals in BCFT. 
Such boundary states are not only conformally invariant, but must satisfy 
some extra conditions. 
Indeed, any linear combination of conformally invariant boundary states
is conformally invariant, whereas the number of physical boundary states are
usually finite. 
One of the most powerful and systematic method for finding such physical
boundary states is Cardy's fusion method\cite{cardy3}, which we shall review 
briefly. 

The extra condition used in Cardy's method is the duality in boundary
partition functions. 
The partition function calculated in the open-string channel and the closed 
string channel leads to different expressions, and their equivalence gives a 
constraint on the boundary states.
In the open string channel, the partition function is a sum of the
chiral characters, 
$Z_{\tilde\alpha\tilde\beta}(q)=\sum_jn^j_{\tilde\alpha\tilde\beta}\chi_j(q)$,
where $\tilde\alpha$ and $\tilde\beta$ stand for boundary conditions on 
the two ends of an open string, $n^j_{\tilde\alpha\tilde\beta}$ is a
non-negative integer representing the multiplicity,
and $\chi_j(q)$ is the character for the
representation $j$. 
This means $n^j_{\tilde\alpha\tilde\beta}$ copies of 
the representation $j$ appear in the bulk when the conditions of two boundaries
are $\tilde\alpha$ and $\tilde\beta$.
We have introduced the modular parameter $q$ as $q=e^{-\pi L/T}$. 
In the closed string channel, the partition function is a tree-level 
amplitude of a closed string propagating from one boundary $\tilde\alpha$ to 
the other $\tilde\beta$, which is written as
$\langle\tilde\alpha\vert e^{-TH}\vert\tilde\beta\rangle$.
Here, $H$ is the Hamiltonian $H=(2\pi/L)(L_0+\bar L_0-c/12)$.
Using the modular parameter $\tilde q=e^{-4\pi T/L}$ the amplitude becomes
$\langle\tilde\alpha\vert(\tilde q^{1/2})^{L_0+\tilde L_0-c/12}
\vert\tilde\beta\rangle$.
The duality of the partition function now demands
$Z_{\tilde\alpha\tilde\beta}(q)
=\langle\tilde\alpha\vert(\tilde q^{1/2})^{L_0+\tilde L_0-c/12}
\vert\tilde\beta\rangle$, which is called Cardy's consistency condition.
Boundary states satisfying the above condition, which we call {\em consistent}
boundary states and denote with tilde (~$\tilde{}$~), may be expanded 
with a complete set of the space of boundary states $\{\langle a\vert\}$ and
$\{\vert a\rangle\}$. Cardy's condition is now written as
\beq
\sum_jn^j_{\tilde\alpha\tilde\beta}\chi_j(q)
=\sum_{a,b}\langle\tilde\alpha\vert a\rangle
\langle a\vert(\tilde q^{1/2})^{L_0+\tilde L_0-c/12}\vert b\rangle
\langle b\vert\tilde\beta\rangle.
\label{eqn:consistency}
\eeq
By solving this equation, the consistent boundaries are expressed as linear
sums of the basis states $\{\langle a\vert\}$ and $\{\vert a\rangle\}$.
A convenient set of such basis states is the Ishibashi 
states $\vert j\rishi$\cite{ishibashi}, which diagonalise the above closed
string amplitudes and give characters\footnote{Although in some literature
the term `Ishibashi state' is used to mean any boundary state satisfying the
condition (\ref{eqn:Tcondz}), we use this term in a narrower sense meaning
the particular solution found by Ishibashi\cite{ishibashi}.
In this paper we call the states including coherent, Ishibashi and consistent 
boundary states collectively as `boundary states,' whereas some authors use 
this term for what we call `consistent boundary states' here.}:
\beq
\lishi i\vert(\tilde q^{1/2})^{L_0+\tilde L_0-c/12}\vert j\rishi
=\delta_{ij}\chi_j(\tilde q).
\label{eqn:ishiamp}
\eeq
We may, however, choose any set of boundary states for the basis
as long as they are complete.

In order to use the above machinery and express the consistent boundary states
in terms of the coherent states we found in the last section, we need
to calculate the closed string amplitudes between
$\langle B(\alpha)\vert$ and $\vert B(\beta)\rangle$. 
Such amplitudes generally involve screening operators, or floating charges
in the bulk. 
Let us consider the situation where $m$ positive ($\alpha_+$) and $n$ negative 
($\alpha_-$) floating charges are present.
The closed-string amplitude for such a process is
\bea
{\cal A}_{\alpha,\beta}
&=&\langle B(\alpha)\vert e^{-TH}Q_+^m Q_-^n\bar Q_+^m\bar Q_-^n
\vert B(\beta)\rangle\nonumber\\
&=&\langle B(\alpha)\vert(\tilde q^{1/2})^{L_0+\bar L_0-c/12}
Q_+^m Q_-^n\bar Q_+^m\bar Q_-^n
\vert B(\beta)\rangle,\nonumber\\
\label{eqn:amp1}
\eea
where $Q_\pm$ is defined in (\ref{eqn:screening}) and 
\bea
&&\bar Q_\pm\equiv\oint d\bar z\bar V_\pm(\bar z),
\label{eqn:barscreening}\\
&&\bar V_\pm(\bar z)=:e^{i\sqrt 2\alpha_\pm\bar\varphi(\bar z)}:.
\label{eqn:barvpm}
\eea
The integration contours must be non-self-intersecting closed curves with
non-trivial homotopy.
In our geometry such contours are the ones which simply go around the cylinder
just once.
A comment on the uniqueness of the amplitude (\ref{eqn:amp1}) is in
order.
It is easy to show that
$[Q_+,Q_-]=0$, $[\bar Q_+,\bar Q_-]=0$. 
Also, $[Q_\pm,\bar Q_\pm]=0$, $[Q_\pm,\bar Q_\mp]=0$ because the holomorphic
and antiholomorphic mode operators commute.
As the screening operators have trivial conformal dimension, they commute
with the Virasoro operators: $[L_n,Q_\pm]=0$, $[\bar L_n, \bar Q_\pm]=0$.
In particular, $[L_0,Q_\pm]=0$ and $[\bar L_0, \bar Q_\pm]=0$.
Hence the order and the position of the screening operators do not matter
and the amplitude with $m$ positive and $n$ negative floating charges may be 
always written in the form (\ref{eqn:amp1}). 

The numbers of the screening charges $m$ and $n$ are not arbitrary but
they must satisfy the charge neutrality condition (otherwise the amplitude
vanishes).
Note that our formalism (see the normalisation (\ref{eqn:hwvnorm})) demands 
charge neutrality in both holomorphic and antiholomorphic sectors. 
In the holomorphic sector, we have charges $-\alpha$ and $\beta$ on the 
boundaries, and $m$ positive and $n$ negative screening charges in the bulk. 
The total charge in the holomorphic part is then
\beq
-\alpha+\beta+m\alpha_++n\alpha_-,
\label{eqn:holocharge}
\eeq
which must be zero.
Similarly, the total charge in the antiholomorphic part is
$-\bar\alpha+\bar\beta+m\alpha_++n\alpha_-$, or, using the condition 
(\ref{eqn:alpha}),
\beq
\alpha-\beta+m\alpha_++n\alpha_-,
\label{eqn:aholocharge}
\eeq
which is also zero.
Since the sum of the holomorphic and antiholomorphic charges must also vanish,
summing the above two expressions we have
$m\alpha_++n\alpha_-=0$.
Now let us recall that the screening charges of the minimal models are
characterised by two co-prime integers $p$ and $p'$ ($p>p'$) as
$\alpha_+=\sqrt{p/p'}$, $\alpha_-=-\sqrt{p'/p}$. 
Then we have
\beq
pm-p'n=0.
\label{eqn:neutrality}
\eeq
Since $p$ and $p'$ are co-prime, $m$ and $n$ are written using an integer $l$ 
as
$m=lp'$, $n=lp$. 
This means the net floating charges must vanish in both holomorphic and 
antiholomorphic sectors. 
The simplest charge configuration obeying this condition is $m=n=0$,
or {\em no} screening operators.
In this case the amplitude (\ref{eqn:amp1}) is particularly easily evaluated.
The oscillating part is calculated with the Heisenberg algebras
(\ref{eqn:heisenberg}) (\ref{eqn:anti-heisenberg}) and repeated use of
Hausdorff formula, as
\beq
\prod_{k=1}^{\infty}\frac{1}{1-\tilde q^k}
=\frac{\tilde q^{1/24}}{\eta(\tilde\tau)}.
\eeq
The zero-mode part,
\beq
\langle\alpha,\bar\alpha;\alpha_0\vert(\tilde q^{1/2})^{(a_0^2+\bar a_0^2)/2
-\sqrt 2\alpha_0(a_0+\bar a_0)-c/12}\vert\beta,\bar\beta;\alpha_0\rangle,
\eeq
is simplified with the central charge (\ref{eqn:central}), the condition on
boundary charges for conformal invariance (\ref{eqn:alpha}) and the operation
of zero-modes on the highest-weight vectors 
(\ref{eqn:zeromodeop1})-(\ref{eqn:zeromodeop4}), as
\beq
\langle\alpha,2\alpha_0-\alpha;\alpha_0\vert 
\tilde q^{\alpha^2-2\alpha_0\alpha+\alpha_0^2-1/24}
\vert\beta,2\alpha_0-\beta;\alpha_0\rangle.
\eeq
Using the normalisation of the highest weight vectors (\ref{eqn:hwvnorm}) 
we have 
\bea
{\cal A}_{\alpha,\beta}
&=&\langle B(\alpha)\vert(\tilde q^{1/2})^{L_0+\bar L_0-c/12}
\vert B(\beta)\rangle\nonumber\\
&=&\frac{\tilde q^{(\alpha-\alpha_0)^2}}{\eta(\tilde\tau)}\kappa'
\delta_{\alpha,\beta}.
\label{eqn:amp2}
\eea
Note the similarity of these amplitudes to the characters (\ref{eqn:cbfschar}) 
of CBFS. This is not a coincidence, but is understood as follows.

Just as a BCFT on the upper half plane can be viewed as a chiral CFT on the 
full-plane by the mirroring procedure we mentioned at the beginning of the
last section, a BCFT on the annulus may be viewed in two ways (Fig.1). 
We can see the system as the holomorphic (${\cal H}$) and antiholomorphic 
($\bar{\cal H}$) sectors residing on a finite cylinder, and they are tied 
together on the two boundaries (a).
We can map the antiholomorphic sector $\bar{\cal H}$ to a continuation 
of the holomorphic sector, via a parity transformation ${\cal P}$. 
This introduces a torus with two cells separated by two boundaries (b). 
We may also apply a time-reversal transformation ${\cal T}$ as well as
${\cal P}$, so that one can go from ${\cal H}$ through a boundary to 
the mapped antiholomorphic sector ${\cal TP(H)}$, and then through the other
boundary and back to ${\cal H}$ along one direction of periodic time.
In this way the BCFT on the annulus can be seen as a {\em chiral} theory on 
the torus.
The closed-string picture is based on the non-chiral picture (a), but
the amplitude should also represent the chiral picture on the torus (b).

In order to describe the minimal models, it is convenient to introduce
boundary states $\vert a_{r,s}\rangle$ and $\vert a_{r,-s}\rangle$ defined as 
\bea
&&\vert a_{r,s}\rangle
=\sum_{k\in Z}\vert B(k\sqrt{pp'}+\alpha_{r,s})\rangle,
\label{eqn:mincoh1}\\
&&\vert a_{r,-s}\rangle
=\sum_{k\in Z}\vert B(k\sqrt{pp'}+\alpha_{r,-s})\rangle.
\label{eqn:mincoh2}
\eea
Similarly we define
\bea
&&\langle a_{r,s}\vert
=\sum_{k\in Z}\langle B(k\sqrt{pp'}+\alpha_{r,s})\vert,
\label{eqn:mincoh3}\\
&&\langle a_{r,-s}\vert
=\sum_{k\in Z}\langle B(k\sqrt{pp'}+\alpha_{r,-s})\vert.
\label{eqn:mincoh4}
\eea
These are linear sums of countably many coherent states (\ref{eqn:cohbsket})
and (\ref{eqn:cohbsbra}) defined in the previous section.
Using (\ref{eqn:amp2}) it is shown that
\bea
\langle a_{r,s}\vert&&(\tilde q^{1/2})^{L_0+\bar L_0-c/12}
\vert a_{r',s'}\rangle\nonumber\\
&&=\frac{\Theta_{pr-p's,pp'}(\tilde\tau)}{\eta(\tilde\tau)}
\kappa'\delta_{r,r'}\delta_{s,s'}\nonumber\\
&&=\frac{\Theta_{pr-p's,pp'}(\tilde\tau)}{\eta(\tilde\tau)}
\delta_{r,r'}\delta_{s,s'},
\label{eqn:minamp1}\\
\langle a_{r,-s}\vert&&(\tilde q^{1/2})^{L_0+\bar L_0-c/12}
\vert a_{r',-s'}\rangle\nonumber\\
&&=\frac{\Theta_{pr+p's,pp'}(\tilde\tau)}{\eta(\tilde\tau)}
\kappa'\delta_{r,r'}\delta_{s,s'}\nonumber\\
&&=-\frac{\Theta_{pr+p's,pp'}(\tilde\tau)}{\eta(\tilde\tau)}
\delta_{r,r'}\delta_{s,s'},
\label{eqn:minamp2}
\eea
and
\beq
\langle a_{r,\pm s}\vert(\tilde q^{1/2})^{L_0+\bar L_0-c/12}
\vert a_{r',\mp s'}\rangle
=0.
\label{eqn:minamp3}
\eeq
Here, we have assumed $1\leq r,r'<p'$ and $1\leq s,s'<p$.
See App. A for our convention of Jacobi theta functions.
We have set $\kappa'=1$ in (\ref{eqn:minamp1}) and $\kappa'=-1$ in
(\ref{eqn:minamp2}). 
This means the states $\vert a_{r,s}\rangle$, $\langle a_{r,s}\vert$ belong
to a unitary sector whereas $\vert a_{r,-s}\rangle$, $\langle a_{r,-s}\vert$
belong to a non-unitary sector.
The amplitudes include all the theta functions appearing in the 
characters of minimal models (\ref{eqn:char}) and thus we have reproduced the 
necessary set of boundary states covering the right hand side of the Cardy's 
consistency condition (\ref{eqn:consistency}).
We shall see this in detail for the Ising model in the next section.
It can be easily checked by using (\ref{eqn:minamp1}) - (\ref{eqn:minamp3}) and
the character formula (\ref{eqn:char}) that the states defined as sums of the 
coherent states,
\bea
&&\vert (r,s)\rishi = \vert a_{r,s}\rangle+\vert a_{r,-s}\rangle,
\label{eqn:ishiket}\\
&&\lishi (r,s)\vert = \langle a_{r,s}\vert+\langle a_{r,-s}\vert,
\label{eqn:ishibra}
\eea
diagonalise the amplitude and reproduce the minimal characters.
These states $\vert (r,s)\rishi$ may then be regarded as the Ishibashi states.

Before discussing the Ising model, we have three points to make about the 
boundary states $\{\vert a_{r,s}\rangle, \vert a_{r,-s}\rangle\}$.
Firstly, the amplitudes (\ref{eqn:minamp1}), (\ref{eqn:minamp2}),
(\ref{eqn:minamp3}) are diagonal, i.e. the boundary states are all orthogonal 
to each other.
This is a consequence of the diagonal amplitude (\ref{eqn:amp2}). 
Indeed, since the boundary charges $k\sqrt{pp'}+\alpha_{r,\pm s}$ are all
different for each set of $(r,\pm s,k)$ and the boundary states
$\{\vert a_{r,s}\rangle, \vert a_{r,-s}\rangle\}$ contain no charges in common,
the amplitudes (\ref{eqn:minamp1}), (\ref{eqn:minamp2}) must vanish unless 
$(r,s)=(r',s')$.
The second point is that these boundary states are unique (besides the
degeneracy $(r,s)\leftrightarrow (p'-r,p-s)$) as long as we want 
to reproduce the theta functions as amplitudes between such boundaries.
The infinite sum expressions (\ref{eqn:thetasum}) for the theta functions are 
power series of $q$, and the power is related to the
boundary charge through the expression (\ref{eqn:amp2}). 
By superimposing the boundary charges appearing in the expression of theta
functions, the boundary states are constructed without ambiguity.
Third, the negative-norm states $\vert a_{r,-s}\rangle$ seem to be unavoidable
even for the unitary minimal models. 
The highest-weight vector $\vert\alpha,\bar\alpha;\alpha_0\rangle$ is built 
on the vacuum $\vert 0,0;\alpha_0\rangle$ by operating with 
$e^{i\sqrt 2\alpha\varphi_0}$ and $e^{i\sqrt 2\bar\alpha\bar\varphi_0}$, and
its norm $\kappa'$ is due to the normalisation of the vacuum 
$\langle 0,0;\alpha_0\vert 0,0;\alpha_0\rangle=\kappa'$.
This $\kappa'$ may be rescaled to an arbitrary real number as long as it is 
either positive or negative definite, but the sign cannot be changed by the 
rescaling. 
The states with $\kappa'=1$ and $\kappa'=-1$ (that is, $\vert a_{r,s}\rangle$
and $\vert a_{r,-s}\rangle$ above) therefore belong to different sectors with 
no intersection.

\begin{figure}
(a) Non-chiral picture ~~~~~~~~~~
(b) Chiral picture ~~~
\epsfxsize=45mm
\epsfysize=45mm
\epsffile{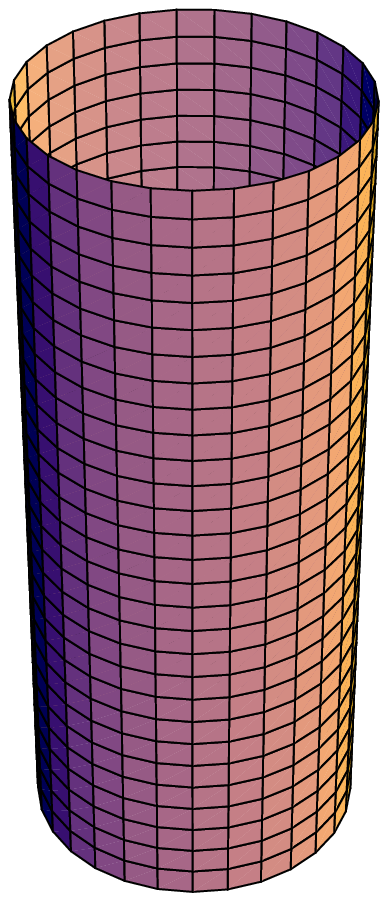}
\epsfxsize=45mm
\epsfysize=45mm
\epsffile{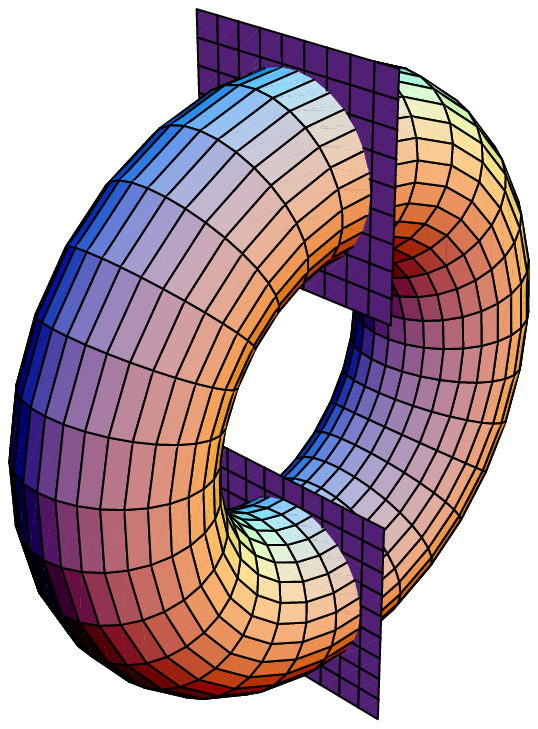}
\caption{BCFT of holomorphic and antiholomorphic sectors glued together on the 
two boundaries of a finite cylinder (a) is equivalently described as a {\em
chiral} theory on a torus (b), where the antiholomorphic sector is now regarded
as continuation of the holomorphic sector, with parity and 
time-reversal transformations. }
\end{figure}


\section{Ising model boundary states}

The Ising model is the simplest non-trivial minimal model, with the two 
characterising co-prime integers $p=4$, $p'=3$, and the central charge $1/2$. 
Also it is clearly one of the most extensively studied critical systems
described by two-dimensional CFT. 
For its detailed description we refer the readers to e.g. \cite{dFMS}.
In this section we demonstrate that the boundary states constructed in the
previous section are enough to reproduce the known physical boundary states
of the Ising model, by a parallel discussion with Cardy's original paper 
\cite{cardy3}. 

The critical Ising model is known to have three physical boundary states, 
corresponding to the two fixed (up and down) and one free boundary
conditions.
They are identified and expressed as particular linear combinations of the 
Ishibashi states by solving the consistency equation 
(\ref{eqn:consistency})\cite{cardy3}. 
The characters of minimal models are linearly transformed under 
$\tau\rightarrow\tilde\tau=-1/\tau$ as
$\chi_i(q)=\sum_j S_{ij}\chi_j(\tilde q)$. 
Substituting this and (\ref{eqn:ishiamp}) into (\ref{eqn:consistency}),
and equating the coefficients of the character functions we have
\beq
\sum_in^i_{\tilde\alpha\tilde\beta}S_{ij}=\langle\tilde\alpha\vert j\rishi
\lishi j\vert\tilde\beta\rangle.
\label{eqn:duality}
\eeq
We assume a state $\vert\tilde 0\rangle$ satisfies the condition
$n^i_{\tilde 0\tilde\alpha}
=n^i_{\tilde\alpha\tilde 0}=\delta^i_{\tilde\alpha}$.
Putting $\tilde\alpha=\tilde\beta=\tilde 0$ in (\ref{eqn:duality}) we have
$\langle\tilde 0\vert j\rishi=\sqrt{S_{0j}}$. 
Similarly letting $\tilde\alpha\neq 0$ and $\tilde\beta=\tilde 0$ we find
$\langle\tilde\alpha\vert j\rishi=S_{\alpha j}/\sqrt{S_{0j}}$.
Substituting these back into (\ref{eqn:duality}) we have
\beq
\sum_in^i_{\tilde\alpha\tilde\beta}S_{ij}
=\frac{S_{\alpha j}S_{\beta j}}{S_{0 j}}.
\label{eqn:verlinde}
\eeq
This is identical to the Verlinde formula\cite{verlinde} and therefore 
$n^i_{\tilde\alpha\tilde\beta}$ is concluded to be the same as the fusion
coefficients.
The consistent boundary states are now expressed using the Ishibashi states as
\beq
\vert\tilde\alpha\rangle
=\sum_j\vert j\rishi\lishi j\vert\tilde\alpha\rangle
=\sum_j\frac{S_{\alpha j}}{\sqrt{S_{0j}}}\vert j\rishi.
\label{eqn:consistentbs}
\eeq
The operators appearing in the Kac table of the Ising model are 
$\phi_{(1,1)}=\phi_{(2,3)}$, $\phi_{(2,1)}=\phi_{(1,3)}$ and
$\phi_{(1,2)}=\phi_{(2,2)}$, which are identified as the identity $I$,
the energy $\epsilon$ and the spin $\sigma$ operators, having the conformal
dimensions $0$, $1/2$, $1/16$, respectively.
Since we know the modular matrix $S_{ij}$ for these representations,
from (\ref{eqn:consistentbs}) we immediately have
\bea
\vert\tilde I\rangle = \vert\tilde 0\rangle
&=&2^{-1/2}\vert I\rishi+2^{-1/2}\vert\epsilon\rishi+2^{-1/4}\vert\sigma\rishi,
\label{eqn:fixed1}\\
\vert\tilde\epsilon\rangle
&=&2^{-1/2}\vert I\rishi+2^{-1/2}\vert\epsilon\rishi-2^{-1/4}\vert\sigma\rishi,
\label{eqn:fixed2}\\
\vert\tilde\sigma\rangle
&=&\vert I\rishi-\vert\epsilon\rishi.
\label{eqn:free}
\eea
Since the first two lines differ only by the sign of the Ishibashi state
$\vert\sigma\rishi$ associated to the spin operator, they are identified as 
the fixed (up or down) boundary states. 
The last line then corresponds to the free boundary state.  

Now let us show that the above procedure can be reproduced using the coherent
states on CBFS instead of the Ishibashi states.
The three characters for the three operators of the Ising model follow
immediately from (\ref{eqn:char}) as
\bea
\chi_I(q)=\chi_{1,1}(q)&=&\frac{1}{\eta(\tau)}
\left[\Theta_{1,12}(\tau)-\Theta_{7,12}(\tau)\right],\\
\chi_\epsilon(q)=\chi_{2,1}(q)&=&\frac{1}{\eta(\tau)}
\left[\Theta_{5,12}(\tau)-\Theta_{11,12}(\tau)\right],\\
\chi_\sigma(q)=\chi_{2,2}(q)&=&\frac{1}{\eta(\tau)}
\left[\Theta_{2,12}(\tau)-\Theta_{10,12}(\tau)\right].
\eea
Using the modular transformation formula of the theta functions 
(\ref{eqn:modularS}) they are written as
\bea
\chi_I(q)&=&\frac{
\Theta_{1,12}(\tilde\tau)+\Theta_{5,12}(\tilde\tau)
-\Theta_{7,12}(\tilde\tau)-\Theta_{11,12}(\tilde\tau)}
{2\eta(\tilde\tau)}\nonumber\\
&&+\frac{
\Theta_{2,12}(\tilde\tau)-\Theta_{10,12}(\tilde\tau)}
{\sqrt 2\eta(\tilde\tau)},\\
\chi_\epsilon(q)&=&\frac{
\Theta_{1,12}(\tilde\tau)+\Theta_{5,12}(\tilde\tau)
-\Theta_{7,12}(\tilde\tau)-\Theta_{11,12}(\tilde\tau)}
{2\eta(\tilde\tau)}\nonumber\\
&&-\frac{
\Theta_{2,12}(\tilde\tau)-\Theta_{10,12}(\tilde\tau)}
{\sqrt 2\eta(\tilde\tau)},\\
\chi_\sigma(q)&=&\frac{
\Theta_{1,12}(\tilde\tau)-\Theta_{5,12}(\tilde\tau)
-\Theta_{7,12}(\tilde\tau)+\Theta_{11,12}(\tilde\tau)}
{\sqrt 2\eta(\tilde\tau)}.\nonumber\\
\eea
These are the character functions appearing in the open-string channel
(left hand side) of the consistency equation (\ref{eqn:consistency}). 
In the closed string channel of (\ref{eqn:consistency}) we expand the
states with $\vert a_{r,\pm s}\rangle$ and $\langle a_{r,\pm s}\vert$
defined in (\ref{eqn:mincoh1}) - (\ref{eqn:mincoh4}), with 
$1\leq r\leq 2$, $1\leq s\leq 3$, and $3s<4r$.
In the Ising model the non-trivial amplitudes (\ref{eqn:minamp1}), 
(\ref{eqn:minamp2}) are
\bea
\langle a_{1,1}\vert (\tilde q^{1/2})^{L_0+\bar L_0-c/12}\vert a_{1,1}\rangle
&=&\Theta_{1,12}(\tilde\tau)/\eta(\tilde\tau),\\
\langle a_{2,2}\vert (\tilde q^{1/2})^{L_0+\bar L_0-c/12}\vert a_{2,2}\rangle
&=&\Theta_{2,12}(\tilde\tau)/\eta(\tilde\tau),\\
\langle a_{2,1}\vert (\tilde q^{1/2})^{L_0+\bar L_0-c/12}\vert a_{2,1}\rangle
&=&\Theta_{5,12}(\tilde\tau)/\eta(\tilde\tau),\\
\langle a_{1,-1}\vert (\tilde q^{1/2})^{L_0+\bar L_0-c/12}\vert a_{1,-1}\rangle
&=&-\Theta_{7,12}(\tilde\tau)/\eta(\tilde\tau),\\
\langle a_{2,-2}\vert (\tilde q^{1/2})^{L_0+\bar L_0-c/12}\vert a_{2,-2}\rangle
&=&-\Theta_{10,12}(\tilde\tau)/\eta(\tilde\tau),\\
\langle a_{2,-1}\vert (\tilde q^{1/2})^{L_0+\bar L_0-c/12}\vert a_{2,-1}\rangle
&=&-\Theta_{11,12}(\tilde\tau)/\eta(\tilde\tau).
\eea
Substituting these into the right hand side of (\ref{eqn:consistency}) and
equating the coefficients of 
$\Theta_{1,12}(\tilde\tau)/\eta(\tilde\tau)$,
$\Theta_{2,12}(\tilde\tau)/\eta(\tilde\tau)$, 
$\Theta_{5,12}(\tilde\tau)/\eta(\tilde\tau)$,
$\Theta_{7,12}(\tilde\tau)/\eta(\tilde\tau)$, 
$\Theta_{10,12}(\tilde\tau)/\eta(\tilde\tau)$ and
$\Theta_{11,12}(\tilde\tau)/\eta(\tilde\tau)$ 
on both sides, we have
\bea
&&\frac 12 n^I_{\tilde\alpha\tilde\beta}
+\frac 12 n^\epsilon_{\tilde\alpha\tilde\beta}
+\frac{1}{\sqrt 2} n^\sigma_{\tilde\alpha\tilde\beta}
=\langle\tilde\alpha\vert a_{1,1}\rangle
\langle a_{1,1}\vert\tilde\beta\rangle,
\label{eqn:coeffs1}\\
&&\frac{1}{\sqrt 2} n^I_{\tilde\alpha\tilde\beta}
-\frac{1}{\sqrt 2} n^\epsilon_{\tilde\alpha\tilde\beta}
=\langle\tilde\alpha\vert a_{1,2}\rangle
\langle a_{1,2}\vert\tilde\beta\rangle,
\label{eqn:coeffs2}\\
&&\frac 12 n^I_{\tilde\alpha\tilde\beta}
+\frac 12 n^\epsilon_{\tilde\alpha\tilde\beta}
-\frac{1}{\sqrt 2} n^\sigma_{\tilde\alpha\tilde\beta}
=\langle\tilde\alpha\vert a_{1,3}\rangle
\langle a_{1,3}\vert\tilde\beta\rangle,
\label{eqn:coeffs3}\\
&&\frac 12 n^I_{\tilde\alpha\tilde\beta}
+\frac 12 n^\epsilon_{\tilde\alpha\tilde\beta}
+\frac{1}{\sqrt 2} n^\sigma_{\tilde\alpha\tilde\beta}
=\langle\tilde\alpha\vert a_{1,-1}\rangle
\langle a_{1,-1}\vert\tilde\beta\rangle,
\label{eqn:coeffs4}\\
&&\frac{1}{\sqrt 2} n^I_{\tilde\alpha\tilde\beta}
-\frac{1}{\sqrt 2} n^\epsilon_{\tilde\alpha\tilde\beta}
=\langle\tilde\alpha\vert a_{1,-2}\rangle
\langle a_{1,-2}\vert\tilde\beta\rangle,
\label{eqn:coeffs5}\\
&&\frac 12 n^I_{\tilde\alpha\tilde\beta}
+\frac 12 n^\epsilon_{\tilde\alpha\tilde\beta}
-\frac{1}{\sqrt 2}n^\sigma_{\tilde\alpha\tilde\beta}
=\langle\tilde\alpha\vert a_{1,-3}\rangle
\langle a_{1,-3}\vert\tilde\beta\rangle.
\label{eqn:coeffs6}
\eea
Let us find the coefficients assuming that they are real
and $\langle\tilde\alpha\vert a_{r,\pm s}\rangle
=\langle a_{r,\pm s}\vert\tilde\alpha\rangle$.
We start by letting $\tilde\alpha=\tilde\beta=\tilde 0$. 
The first equation (\ref{eqn:coeffs1}) gives 
$\vert\langle\tilde 0\vert a_{1,1}\rangle\vert^2=1/2$ and we can choose
$\langle\tilde 0\vert a_{1,1}\rangle=1/\sqrt 2$. Likewise, 
from (\ref{eqn:coeffs2}) - (\ref{eqn:coeffs6}) we find
$\langle\tilde 0\vert a_{2,2}\rangle=\langle\tilde 0\vert a_{2,-2}\rangle
=2^{-1/4}$, 
$\langle\tilde 0\vert a_{2,1}\rangle=\langle\tilde 0\vert a_{1,-1}\rangle
=\langle\tilde 0\vert a_{2,-1}\rangle=1/\sqrt 2$.
Next, letting $\tilde\alpha=\tilde\epsilon$ and $\tilde\beta=\tilde 0$ we find
$\langle\tilde\epsilon\vert a_{1,1}\rangle
=\langle\tilde\epsilon\vert a_{2,1}\rangle
=\langle\tilde\epsilon\vert a_{1,-1}\rangle
=\langle\tilde\epsilon\vert a_{2,-1}\rangle
=1/\sqrt 2$, and
$\langle\tilde\epsilon\vert a_{2,2}\rangle
=\langle\tilde\epsilon\vert a_{2,-2}\rangle=-2^{-1/4}$.
Lastly, putting $\tilde\alpha=\tilde\sigma$ and $\tilde\beta=\tilde 0$ we find
$\langle\tilde\sigma\vert a_{1,1}\rangle
=\langle\tilde\sigma\vert a_{1,-1}\rangle=1$,
$\langle\tilde\sigma\vert a_{2,2}\rangle
=\langle\tilde\sigma\vert a_{2,-2}\rangle=0$
and
$\langle\tilde\sigma\vert a_{2,1}\rangle
=\langle\tilde\sigma\vert a_{2,-1}\rangle
=-1$.
Then the consistent boundary states are expressed in terms of the coherent 
states as
\bea
\vert\tilde I\rangle = \vert\tilde 0\rangle
&=&2^{-1/2}(\vert a_{1,1}\rangle+\vert a_{1,-1}\rangle
+\vert a_{2,1}\rangle+\vert a_{2,-1}\rangle)\nonumber\\
&&+2^{-1/4}(\vert a_{2,2}\rangle+\vert a_{2,-2}\rangle),
\label{eqn:cfixed1}\\
\vert\tilde\epsilon\rangle
&=&2^{-1/2}(\vert a_{1,1}\rangle+\vert a_{1,-1}\rangle
+\vert a_{2,1}\rangle+\vert a_{2,-1}\rangle)\nonumber\\
&&-2^{-1/4}(\vert a_{2,2}\rangle+\vert a_{2,-2}\rangle),
\label{eqn:cfixed2}\\
\vert\tilde\sigma\rangle
&=&\vert a_{1,1}\rangle+\vert a_{1,-1}\rangle
-\vert a_{2,1}\rangle-\vert a_{2,-1}\rangle.
\label{eqn:cfree}
\eea
These are exactly the same result as (\ref{eqn:fixed1}) - (\ref{eqn:free}), 
as the relation between the Ishibashi states and the coherent states are
given in (\ref{eqn:ishiket}) and (\ref{eqn:ishibra}). 
We have thus shown for the Ising model that the coherent states constructed on 
CBFS are not merely a subspace of the boundary states but they cover the space 
spanned by Cardy's consistent boundary states.

In the case of the Ising model, a similar construction of the boundary states
from coherent states has been done using free Majorana 
fermions\cite{leclair,yamaguchi,nepo}. 
In a sense the present analysis is a generalisation of such a construction
to general minimal theories. 


\section{Discussion}

In this paper we have constructed a set of coherent states on CBFS which
preserve the conformal invariance, and argued that Cardy's consistent 
boundary states for minimal models are expressed as linear combinations of 
such states.
We have demonstrated this explicitly in the example of Ising model.
Our approach provides a new intuitive picture of boundary states in CFT, 
in terms of the boundary charges which obey the charge neutrality conditions 
with bulk screening operators.

We would like to conclude this paper by stressing that, apart from giving a 
new interpretation of boundary states, this approach is quite powerful 
in at least two respects.

Firstly, once consistent boundary states are expressed in terms of the coherent
states, it is in principle possible to compute boundary $n$-point functions 
on an annulus directly without resorting to extra information on the boundary. 
The $n$-point function on the upper half plane involving an operator 
$\phi_{r,s}$ is found in the conventional method by solving the 
$(r\times s)$-th order differential equations satisfied by the $2n$-point 
function on the full plane\cite{cardy1}.
Solutions to such a differential equation are in the form 
$A_1F_1+A_2F_2+\cdots$ 
where $F_i$ represent the conformal blocks, and the coefficients $A_i$ 
reflect boundary conditions and are determined by considering e.g. the
asymptotic behaviour of the $n$-point function.
In our Coulomb-gas approach, $n$-point functions on an annulus are obtained 
by inserting vertex operators between the boundary-to-boundary amplitudes, 
with appropriate inclusion of screening operators, leading to an integral
representation of the correlation functions.
In practice, however, such expressions involving multiple integrals of 
theta-functions are not always easy to evaluate. 

The second advantage of our approach is its wide applicability. 
The coulomb-gas approach is not constrained to the minimal models, but it
also applies to WZNW models\cite{gmoms} and CFTs involving 
$W$-algebras\cite{fl,bilal}.
Although we only presented the results of the simplest minimal model here, 
generalizations of our approach to these models also seem to be possible.
We hope to come back to these issues in future publications.


\acknowledgements

I am grateful to John Wheater for fruitful discussions and careful reading 
of the manuscript.
I also appreciate helpful conversations with Ian Kogan and Alexei Tsvelik.
\appendix


\section{Jacobi theta functions}

In this appendix we list some formulae of elliptic functions used in the
main text.
The Jacobi theta function $\Theta_{\lambda,\mu}(\tau)$ and the
Dedekind eta function $\eta(\tau)$ are defined as
\bea
&&\Theta_{\lambda,\mu}(\tau)=\sum_{k\in Z}q^{(2\mu k+\lambda)^2/4\mu},
\label{eqn:thetasum}\\
&&\eta(\tau)=q^{1/24}\prod_{n\geq1}(1-q^n),
\eea
where $q=e^{2\pi i\tau}$.
From this definition it is obvious that $\Theta_{\lambda,\mu}(\tau)$ has the
following symmetries:
\beq
\Theta_{\lambda,\mu}(\tau)=\Theta_{\lambda+2\mu,\mu}(\tau)
=\Theta_{-\lambda,\mu}(\tau).
\eeq
They transform under the modular S transformation ($\tau\rightarrow -1/\tau$)
and the modular T transformation ($\tau\rightarrow\tau+1$) as
\bea
\Theta_{\lambda,\mu}(-1/\tau)
&=&\sqrt{\frac{-i\tau}{2\mu}}\sum_{\nu=0}^{2\mu-1}e^{\lambda\nu\pi i/\mu}
\Theta_{\nu,\mu}(\tau),\nonumber\\
\eta(-1/\tau)&=&\sqrt{-i\tau}\eta(\tau),
\label{eqn:modularS}
\eea
and
\bea
\Theta_{\lambda,\mu}(\tau+1)
&=&e^{\lambda^2\pi i/2\mu}\Theta_{\lambda,\mu}(\tau),\nonumber\\
\eta(\tau+1)&=&e^{\pi i/12}\eta(\tau).
\label{eqn:modularT}
\eea
In the main text we only used the modular S transformation.

The theta functions we used for the Ising model are related to another 
commonly used notation,
\bea
&&\theta_2(\tau)=\sum_{k\in Z}q^{(k+1/2)^2/2},\\
&&\theta_3(\tau)=\sum_{k\in Z}q^{k^2/2},\\
&&\theta_4(\tau)=\sum_{k\in Z}(-1)^kq^{k^2/2},
\eea
by 
\bea
\sqrt{\eta(\tau)\theta_2(\tau)/2}&=&\Theta_{2,12}(\tau)-\Theta_{10,12}(\tau),\\
\sqrt{\eta(\tau)\theta_3(\tau)}
&=&\Theta_{1,12}(\tau)+\Theta_{5,12}(\tau)\nonumber\\
&&-\Theta_{7,12}(\tau)-\Theta_{11,12}(\tau),\\
\sqrt{\eta(\tau)\theta_4(\tau)}
&=&\Theta_{1,12}(\tau)-\Theta_{5,12}(\tau)\nonumber\\
&&-\Theta_{7,12}(\tau)+\Theta_{11,12}(\tau).
\eea

 
\end{document}